# Fine structure of swift heavy ion track in rutile TiO$_2$

Pengfei Zhai[a,b,*], Shuai Nan[c,b], Lijun Xu[a,b], Weixing Li[c,b], Zongzhen Li[a,b], Peipei Hu[a,b], Jian Zeng[a,b], Shengxia Zhang[a], Youmei Sun[a,b], Jie Liu[a,b,†]

[a] *Institute of Modern Physics, Chinese Academy of Sciences, Lanzhou 730000, China*
[b] *University of Chinese Academy of Sciences, Beijing 100049, China*
[c] *CAS Center for Excellence in Tibetan Plateau Earth Sciences, and Key Laboratory of Continental Collision and Plateau Uplift, Institute of Tibetan Plateau Research, Chinese Academy of Sciences, Beijing 100101, China*

**Abstract**

We report on the first observation of fine structure of latent tracks in rutile TiO$_2$, which changes from cylinder to dumbbell-shape and then to sandglass-shape as a function of the ion path length. Based on inelastic thermal spike model, we show that Hagen-Poiseuille flow of molten phase produces the hillocks on surface and the void-rich zone near surface after epitaxial recrystallization due to material deficit, while at a deep depth, the lack of efficient outflow and recrystallization result in the absence of tracks. We propose that "core-shell" duration of transient molten phase induced by swift heavy ion and parabolic distribution of fluid velocity are radial-dependent. Moreover, the various morphologies of tracks are a consequence of the molten phase outflow and recrystallization during rapid cooling down. Our perspective provides a new interpretation in the track formation.

**Keywords**: Ion track; TEM; hillock; recrystallization; Hagen-Poiseuille flow

**1. Introduction**

Rutile (TiO$_2$) polymorphs have widespread applications in solar cell [1,2], photocatalyst [3-5], single-phase ceramic nuclear waste form [6,7] and high dielectric gate insulator for the new generation of metal-oxide-semiconductor field-effect

---

transistor (MOSFET) [8]. Ion irradiations in rutile are of importance because the physical, electronic and photocatalytic performance in $TiO_2$ can be significantly improved [9-15]. As an example, ion implantation has been demonstrated to shift the absorption band to the visible light region in $TiO_2$ [9-13], whereas the wide bandgap in the conventional $TiO_2$ results in only about 3-5% of the total solar light for energy conversion. However, the failure of material used in nuclear waste form induced by fission fragment [6,7] or breakdown of gate oxide in MOSFET caused by space irradiation [16] is harmful to the service life. Therefore, it is a long-standing objective in material science to understand the correlation between intrinsic properties and its behavior to energetic heavy ions.

Rutile $TiO_2$ irradiated by swift heavy ions has been investigated intensely by transmission electron microscopy (TEM) [17-19], atomic force microscopy (AFM) [20-24], Rutherford backscattering in channeling mode (RBS/C) [23], and grazing-incidence small-angle X-ray scattering (GISAXS) [24]. Hillocks on the surface of specimen, after bombarded with swift heavy ions under normal or grazing incidence angle, were observed by AFM and even scanning electron microscopy [20-24]. The response of $TiO_2$ polymorphs to heavy ions irradiation were investigated by molecular dynamics (MD) simulations [25]. It was shown that rutile $TiO_2$ has strong recrystallization ability [25]. More recently, the conical tracks consist of an agglomeration of 1–2 nm sized features near surface of rutile, was observed by TEM [17,18]. As the detailed morphology and structure of irradiation-induced damage along the entire length of ion tracks has not been revealed, the reduced internal pressure near surface of rutile has been considered responsible for the formation of conical morphology of ion track [17].

In this study, by applying oblique incidence, we are able to observe the morphological change at each place along the entire length of an ion track in ion-irradiated rutile, which consists of a pair of hillocks on surfaces, porous ion track beneath the hillocks, and undamaged regime below the tracks. Furthermore, to our knowledge, we first observed the track morphology changing from cylinder to dumbbell-shape and then to sandglass-shape as a function of the ion path length. This full picture of the ion track allows us to systematically investigate the formation and structure of both damaged and undamaged regimes. Based on inelastic thermal spike (i-TS) model, we give a new insight into the ion track formation.

## 2. Experimental

### 2.1 Samples and swift heavy ion irradiation

Single crystals of rutile $TiO_2$ with sample size $10 \times 10 \times 0.5$ mm$^3$ or $5 \times 5 \times 0.5$ mm$^3$ and two different orientations of (110) and (001) were purchased from KJ-MTI Corporation, China. Surfaces were epi-polished down to root mean square roughness $\leq 0.2$ nm and the crystal miscut angle is less than $0.5^o$. The lamella samples were prepared by crushing in a mortar, mixing with ethanol, and depositing on a copper TEM grid covered with an amorphous carbon thin film.

The samples were irradiated with initial kinetic energy of 19.5 MeV/u $^{129}$Xe at a fluence of $5 \times 10^{10}$ ions/cm$^2$ for $0^o$ incidence (with respect to the normal of surface), $2 \times 10^{10}$ ions/cm$^2$ for $45^o$ incidence, 12.5 MeV/u $^{181}$Ta at $1 \times 10^{10}$-$5 \times 10^{11}$ ions/cm$^2$ for $0^o$, 7° and 45° incidence, 13.5 MeV/u $^{181}$Ta at $3\times10^{10}$ ions/cm$^2$ for 0° incidence and 9.5 MeV/u $^{209}$Bi at $5\times10^{11}$ ions/cm$^2$ for 0° and 60° incidence at Heavy Ion Research Facility in Lanzhou (HIRFL) at room temperature. The ion flux detected on-line by a detector with three aluminum foils (total thickness 18 μm) was less than $2\times10^8$ ions/(cm$^2$ s) to avoid macroscopic sample heating. The fluence was determined with an estimated uncertainty of 10-20%. Aluminum degraders (99.99% purity) of different thicknesses were usually placed in front of the samples in order to change the energy. The detailed irradiation parameters are listed in Table 1. The calculated ion range in bulk samples is around 17-100 μm. However, by changing incidence angle, the actual penetrating length in a TEM sample was around 50-150 nm, which enables us to systematically investigate the change in track morphology as a function of ion length.

**2.2 AFM, TEM and Raman characterization**

To investigate the topography of the surface after irradiation, tapping mode AFM was performed using Cypher (Asylum Research, USA) at ambient atmosphere and Arrow-UHF (NanoWorld, Switzerland) with cantilever resonance frequencies of about 1.4 MHz. The images were analyzed using the built-in analysis software. The irradiated lamella samples were observed using a Tecnai G2 F20 S-TWIN TEM (FEI, USA) operating at 200 kV.

Ultraviolet (UV) Raman spectra were measured at room temperature with a LabRAM HR Evolution (Horiba Jobin Yvon, France) with spectral resolution of 2 cm$^{-1}$. The excitation wavelength was 325 nm, the object lens was 74× and the power at the sample was about 2 mW. Visible Raman spectra, excited by 532 nm laser, were obtained at room temperature on a LabRAM HR800 (Horiba Jobin Yvon, France) with spectral resolution of about 0.7 cm$^{-1}$. The object lens was 50× and the power at the sample was below 2 mW. The confocal pinhole was 100 μm. As rutile $TiO_2$ is transparent in the visible light region, the sampled depth of 532 nm is estimated to be 2 μm (the longitudinal spatial resolution is ~ 2 μm), while the penetration depth of 325 nm in $TiO_2$ is about 20 nm due to very strong absorption in ultraviolet region [27].

## 3. Results

### 3.1 Track morphology as a function of ion path length

Figure 1 shows the bright field TEM images of ion tracks in lamella rutile induced by 1390 MeV Bi ions, in which four different track morphologies can be distinguished. In Fig. 1(a), an individual ion track appears as nearly cylindrical and continuous. It consists of irregular structures, which have white core and dark fringe at under focus condition, dark core and white fringe at over focus condition, and little contrast at focus condition. This can be understood by Fresnel contrast, a typical TEM technique to view cavities or highly porous tracks [17,28-32]. At both ends of an ion track, circular features assigned as spherical hillocks formation at the entrance and exit surfaces of ion trajectory [28,29,32] are also visible. It was reported that the hillocks of irradiated rutile $TiO_2$ are crystalline [17], just like the hillocks of irradiated $CeO_2$ [29], because the lattice spacing and orientation of the hillocks coincide with those of the matrix. In this study, the average size of the hillocks (10.2 ± 1.6 nm) is slightly larger than that of the corresponding track (8.4 ± 1.6 nm). Note that several small isolated islands are visible (Fig. 1(a)), with an average diameter of the isolated islands about half of the hillocks. The inhomogeneous contrast inside the tracks can be identified clearly by HRTEM as shown in Fig. 1(b). Except a few faceted structures, the voids appear almost irregular, significantly different from the well aligned, faceted anion voids, which are a sign of preferential creation voids along specific directions, in irradiated $CaF_2$ [28].

In Fig. 1(c), the entire length of a dumbbell-shaped ion track irradiated with Bi ions is visible by TEM under overfocused condition. The "waist" of the ion track is marked by arrows. This dumbbell-shaped morphology is different from the cylindrical morphology shown in Fig. 1(a). It is not due to the reducing of electronic energy loss of 1390 MeV Bi ions in lamella rutile, because the electronic energy loss is almost unchanged (39.9 keV/nm) for such a short depth change in this thin sample. Moreover, the narrow part is just in the middle of the tracks. We interpret this morphology discrepancy in terms of the difference of ion path length. Note that the real ion path length of the two tracks in Fig. 1(c) is longer.

Our assumption that track morphologies are different for different ion path lengths is further confirmed, as an example is shown in Fig. 1(d). In the bottom right corner, the track morphology marked by 1 appears dumbbell-shaped. However, the tracks whose length are much longer than the dumbbell-shaped one, display a new morphology, named as sandglass-shape. A slight difference for the two sandglass-shaped tracks (labeled by 2 and 3) is whether the narrowed part (~1 nm in radius) of the track is continuous. Actually, this small difference is also associated with the ion path length and it will be discussed in more detail in the following text.

These track morphologies were reexamined in lamella samples irradiated by 1703 MeV Ta ions with 45° incidence (Fig. 2). Different morphologies can be visible in this TEM image at the same time (Fig. 2(a)). Thanks to the oblique incidence, the apparent length of ion track observed by TEM has a positive correlation with its actual length in the sample (Fig. 2(b)). It can be found the short track is nearly cylindrical, the longer one is dumbbell-shaped, and the longest one is sandglass-like and discontinuous.

Furthermore, the cross-sectional TEM image of sample prepared by focused ion beam technology from bulk rutile $TiO_2$ irradiated with 1390 MeV Bi ions were shown in Fig. 3. The length of the conical track is about 60 nm and it is consistent with the previous work [17,18].

In Fig. 4, a schematic is used in order to systematically describe the five different morphologies of ion track in rutile as a function of ion path length, based on our TEM observations.

### 3.2 Hillocks on surfaces

The topographies of the irradiated rutile surface characterized by AFM are shown in Fig. 5. The hillocks induced by 1390 MeV Bi ions at a fluence of $5 \times 10^{11}$ ions/cm$^2$ appear almost overlapped (Fig. 5(a)). However, the estimated ratio of ion affected region should be only 39%, considering the diameter of the individual hillock being ~ 10 nm (Fig. 1). It is because the finite curvature radius of the tip (nominally 10 nm for Arrow-UHF tip) induces the notable broadening effect for the hillocks in lateral direction. However, the resolution in vertical direction is mainly affected by the noise of the Z sensor of AFM (<50 pm in closed loop condition for Cypher) instead of the finite curvature radius of the tip. Moreover, the hillock height determined by AFM has been considered as an appropriate parameter to describe the behavior of ion track formation versus the electronic energy loss in $Y_3Fe_5O_{12}$ (YIG) [33]. By assuming that the spherical hillocks exactly protrude out of the specimen surface, as in the cases of the irradiated $CeO_2$ [29], we carefully used the average hillock height to represent the diameter of track, despite that the track diameter is slightly smaller than the spherical diameter of hillock (Fig. 1 and 2). The hillock yields ($\eta$, hillock number/ion fluence) and heights ($D_h$) with standard deviation induced by Xe and Ta ions were analyzed using built-in analysis software through Gaussian fitting to the center heights of the hillocks. Considering the experimental errors (10-20%) in the ion fluence, one can conclude that an incident ion almost creates one hillock as the calculated hillock yield is about 80-100%. The heights of the hillocks are presented in Table 1.

Alternatively, the corrected diameters of hillocks can be also obtained (Table 1) using a simple deconvolution approach $D_c = D_a^2/4D_t$ [34], where $D_c$ is the corrected diameters of hillocks, $D_a$ is the apparent diameter of hillocks and $D_t$ is the tip diameter (20 nm). These corrected diameters are actually very close to the heights of hillocks. It confirms that the hillocks are spherical and the whole sphere almost reaches out of the surface.

**3.3 Visible and UV Raman spectra of irradiated rutile $TiO_2$**

The bulk rutile with orientation of (001) irradiated by 1390 MeV Bi ions at the fluence of $5\times10^{11}$ ions/cm$^2$ was also characterized by Raman spectroscopy as shown in Fig. 6. The typical Raman bands of pristine rutile TiO$_2$ locate at 143, 235, 447 and 612 cm$^{-1}$, which have been assigned to the B$_{1g}$, higher order (HO) band, E$_g$ and A$_{1g}$ modes, respectively [35]. Besides, a band at 826 cm$^{-1}$ ascribed to the B$_{2g}$ mode appears in the UV Raman spectra [36]. For 532 nm excitation, no distinct change can be observed, except for the slight Raman intensity decrease of irradiated sample. In contrast, under 325 nm excitation the Raman spectrum of irradiated rutile displays significant changes: (1) the Raman intensity decrease dramatically and (2) the band B$_{1g}$ at 143 cm$^{-1}$ and the shoulder band of A$_{1g}$ at ~ 700 cm$^{-1}$ disappear. It was reported that the B$_{1g}$ mode is very sensitive to the long-range order of rutile TiO$_2$ and the vanishment of B$_{1g}$ means the loss of long-range ordering in the sampled depth of the irradiated rutile [37]. Also, the B$_{1g}$ mode cannot be observed in nanocrystalline rutile TiO$_2$ [38]. It should be emphasized that the sampled depths for 532 and 325 nm excitation sources in rutile are about 2 μm and 20 nm, respectively. It indicates that the damage near surface of irradiated rutile is much more severe than the bulk. This result is consistent with the above TEM observation. Interestingly, the RBS/C spectra of the irradiated yttria-stabilized cubic ZrO$_2$ (YSZ) by swift heavy ions revealed the similar phenomenon [39]. It was found that the amount of radiation damage is significantly higher in the first 100 nm layer below the surface, while it is almost constant and low in the deep [39].

**4. Discussion**

**4.1 i-TS model calculations versus experimental results**

Based on the inelastic thermal spike (i-TS) model [40-42], two classical heat diffusion equations can be used to calculate the energy exchange between the electronic subsystem and atomic subsystem, driven by the temperature difference ($T_e$ - $T_a$) between the electronic subsystem temperature ($T_e$) and atomic subsystem temperature ($T_a$), which is stimulated by the energy inputs into the electronic subsystem $A(r[\upsilon], t)$ from the electronic stopping powers for specific ion velocity $\upsilon$:

$$C_e(T_e)\frac{\partial T_e}{\partial t} = \frac{1}{r}\frac{\partial}{\partial r}\left[rK_e(T_e)\frac{\partial T_e}{\partial r}\right] - g(T_e - T_a) + A(r[\upsilon], t)$$

$$C_a(T_a)\frac{\partial T_a}{\partial t} = \frac{1}{r}\frac{\partial}{\partial r}\left[rK_a(T_a)\frac{\partial T_a}{\partial r}\right] + g(T_e - T_a)$$

where $C_e$, $C_a$, $K_e$ and $K_a$ are the specific heats and thermal conductivities of the electronic and atomic subsystems, respectively. The electron-phonon coupling strength g is linked to the electron-phonon mean free path $\lambda$ by the relation $\lambda \sim K_e/g^2$, which approximately equals $B/g^2$ (with $B$= 2 J/(s cm K)) in insulators [41,42], and is scaled with the band gap $E_g$ as determined by experimental results [41-44]. Another key parameter in this model, the energy to melt $E_m$ [41-44], is 0.69 eV/at for rutile [45]. The detailed physical properties of rutile $TiO_2$ used for the i-TS model calculation are presented in Table 2.

Using the thermal spike model, Awazu et al. [19] calculated the ion track size directly from the radius of lattice temperature over the equilibrium melting point $T_m$ of rutile (2130 K). However, the latent heat of fusion (838.07 kJ/g [45]), which is crucial for the transition of rutile from solid to liquid, was not considered. Furthermore, due to the transient and huge heating rate of the atoms, $T_m$ is actually not the adequate parameter to characterize the melting process, as experimentally proven by femtosecond laser experiments [49,50]. As was also suggested by Rethfeld et al. [51], the temperature that allows for homogeneous nucleation of a liquid phase induced by ultra-short pulsed laser irradiation should be ~$1.4T_m$ within a few picoseconds. Thus, without considering superheating, one probably underestimates the electronic energy loss threshold of track formation in rutile, and in turn overestimates the size of ion tracks under high electronic excitation [40,41].

We assume that the melting criterion is associated with the ion track formation in irradiated rutile. The calculated results by i-TS model for 1390 MeV Bi irradiation are displayed in Fig 7. Within superheating scenario, the energy to melt $E_m$ has been considered as the criterion for the transition from solid to liquid phase [40,41,52]. For rutile, the energy to reach the melting point is 0.46 eV/at (initial temperature 300 K), while the solid to liquid phase transition energy (latent heat of fusion) is 0.23 eV/at [45]. Thus, the corresponding radius 4.3 nm can be determined by $E_m = 0.69$ eV/at (Fig. 7(a)). It must be emphasized that i-TS model just gives a prediction for initial track size without considering a recrystallization process. One can notice that the lifetime of molten $TiO_2$ for different radial distances from the core is different (Fig. 7(a)). For instance, the duration of molten phase for radius of 0.5 and 4.3 nm are about $t_{0.5}$=11.1 ps and $t_{4.3}$=1.5 ps, respectively. This core-shell distribution due to different duration of melted phase will be used to explain the evolution of track morphology. The superheating atomic temperature versus time is also given in Fig. 7(b). For the radius of 4.3 nm, superheating melting temperature $T_S$ is determined as 2900 K with the heating rate ~$10^{15}$ K/s. The calculated $T_S$ is larger than the equilibrium melting point (2130 K) by a factor of 1.36, consistent with the calculated result by Rethfeld et al. [51], where transient thermal melting was considered as homogeneous nucleation of the liquid phase in a femtosecond laser-excited bulk.

The experimental radii from previous work [18,19,23,53,54] and the present work are compared with i-TS calculations within superheating scenario (Fig. 8). It indicates that most of experimental data agree quite well with i-TS model calculations, with one exception, as marked with the open circle with cross reproduced from Awazu et al. [19]. For example, the track radius induced by 115 MeV Br (16.1 keV/nm) is ~ 2.6 times higher than the prediction of i-TS model. The discrepancy is possibly because the authors considered the etched track as latent ion track. However, after 20% HF etching for 10s, both the stressed part and the amorphous core were etched. According to the Fig. 1(c) from Ref. 19, the unetched track radius induced by 115 MeV Br is only about 1.0 nm, evidently smaller than the reported size of 3.7 nm after etching. Note that in another paper by Awazu et al. [53], the radius of unetched track induced by 84.5 MeV Cu ions (13.6 keV/nm) is $0.9 \pm 0.2$ nm, which is consistent well with our i-TS prediction (open circle). It was reported that after a high fluence ($>6\times10^{12}$ ions/cm$^2$) of 167 MeV Xe irradiation, the rutile crystal was amorphized, extending from the surface down to 8.3 μm below, where the remaining energy is about 15 MeV with a corresponding electronic stopping power ~7.3 keV/nm [18]. This value coincides well with the present i-TS prediction for 0.1 MeV/u ion irradiation (Fig. 8), indicating that the threshold for electronic energy loss is ~7 keV/nm.

**4.2 Why is the deduced track size by RBS/C close to the i-TS model prediction?**

As observed by TEM, the morphology of ion track near the surface of rutile is conical (Fig. 3) and the track radius in the deep is much thinner (~1 nm as shown in Fig. 1(d)). Logically, the effective radius deduced from RBS/C, which assumes that the tracks have identical size and cylindrical symmetry [23], is intuitively expected to be much smaller than our calculation by i-TS model. However, it is interesting that the deduced track size from RBS/C [23] agrees well with the present i-TS model prediction (Fig. 8).

We suggest that it is the strained region [19,53] and/or the defective channel around the track [55,56], which the RBS/C measurement is also sensitive to within its detection depth, attribute to the additional backscattering yield. Although RBS/C cannot obtain the final track size, if knowing that the timescale of track formation is usually several picoseconds, it gives the estimation of initial track size for the irradiated rutile with swift heavy ions.

**4.3 What are responsible for morphology evolution of ion tracks?**

A reduced internal pressure near surface of rutile has been suggested to be responsible for the formation of conical morphology of ion track [17]. However, this model cannot explain why track morphology changes from cylinder to dumbbell-shape or sandglass-like shape as a function of ion path length in nearly the same thickness of thin sample as shown in Fig. 1 and 4. More importantly, as the authors point out, if the reduced internal pressure holds true, these near-surface conical features should be a general phenomenon. However, the ion tracks near surface in apatite [57], YIG [58,59] and $Gd_2TiO_7$ [60] are clearly cylindrical. As a fact, this phenomenon was observed undoubtedly only in rutile $TiO_2$, YSZ and $Al_2O_3$ single crystal so far [17,18,39,61]. The conical track also appeared in irradiated $Gd_2Zr_2O_7$ pyrochlore as shown in Fig.2(f) of Ref. 62, although the authors did not pay attention to this.

We suggest that the near-surface conical feature with small voids is not a general phenomenon, but limited to one kind of materials that have a strong ability to recrystallize according to the matrix structure [25,63-65], and the molten phases have low viscosity and low density. The latter two enable the molten material near surface flow outwards efficiently and cause large mass deficiency for the voids formation in the track.

Specifically, we propose that the morphology evolution of tracks in rutile $TiO_2$ as a function of ion path length is a consequence of the molten phase outflow and recrystallization [25] during rapid quenching. Outflow is mainly related to two factors: lifetime and velocity of the fluid. The specific lifetime of molten region induced by swift heavy ions is scaled with its radial distance from the core. The material closer to the core is able to keep the liquid phase for longer time (Fig. 7). It has been proven that the increased lifetime of molten $TiO_2$ induces bigger hillock [66].

Here, we give a quantitative description in the case of irradiated rutile by 1390 MeV Bi ions. The track size $R_0$ (4.3 nm) and cylindrical morphology in rutile calculated by i-TS model can be regarded as the initial track size and shape. First, the mean fluid velocity $v_{mean}$ can be estimated from the equation (1) - (4), assuming the spherical hillock formed due to the molten $TiO_2$ expelled from the initial track near surface (mass conservation):

$$V_f \cdot \rho_l = V_h \cdot \rho_s \quad (1)$$

$$V_h = \frac{4}{3}\pi R_h^3 \quad (2)$$

$$V_f = \int_0^{R_0} 2\pi r v_{mean} \cdot t(r) dr \quad (3)$$

$$t(r) = 10.02 + 0.01r + 0.11r^2 - 0.13r^3 \quad (4)$$

where, $V_f$ and $V_h$ are the volume of molten $TiO_2$ flowing onto the surface and that of crystalline hillock, respectively, $\rho_l$ is density of molten $TiO_2$ (3.21 g/cm³ [46]), $\rho_s$ is density of crystalline rutile (4.25 g/cm³) and $R_h$ is the radius of hillock (5.1 nm). Duration of molten phase $t(r)$ is a function of radial distance $r$ from the core of thermal spike, which can be fitted by the i-TS calculation. Combining the equation (1) - (4), we get the $v_{mean}$ as 1.82 nm/ps (1820 m/s).

Then, according to the famous Reynold's number ($R_e$) formula in Fluid Mechanics,

$$R_e = \frac{d \cdot \rho_l \cdot v_{mean}}{\mu} \quad (5)$$

where, $d$ is the diameter of initial track (8.6 nm), $\mu$ is the viscosity of molten $TiO_2$ (~ 30 mPa·s for 67-80% $TiO_2$ slags [67]). We can get $R_e$ of the fluid as 1.67, which is much less than 2000. So, the fluid of molten $TiO_2$ is laminar flow.

For the laminar flow through a straight circular tube (so called Hagen-Poiseuille flow), the velocity distribution of the flow is parabolic in profile [68], where the fluid velocity is zero at the wall of the tube (non-slip condition), while the velocity reach the maximum ($v_{max}$) in the center of tube and $v_{max}=2v_{mean}$ [68]. Moreover, the radius-dependent velocity can be shown as equation (6) [68]:

$$v(r) = v_{max}\left(1 - \frac{r^2}{R_0^2}\right) \quad (6)$$

where, $v_{max}$=3.64 nm/ps. Combining Eq. (3), (4) and (6), we can get the volume of molten $TiO_2$ flowing onto the surface as 883 nm$^3$. The mass difference between outflow fluid and spherical hillock ($V_f \cdot \rho_l$ - $V_h \cdot \rho_s$ )/$V_h \cdot \rho_s$ is about 22%.

From equation (6), it shows that fluid velocity of fluid is also radius-dependent. Therefore, in the core region, the longest duration and highest fluid velocity of liquid phase induce the most (longest) outflow of the liquid phase. In contrast, a little material flows out of the surface near the interface of initial liquid region and solid matrix. After recrystallization during the quenching of the liquid region, the region far from the core and surface are recrystallized, just leaving the conical track with many voids near surface of sample. It can be roughly estimated that the mass deficit is ~15% (i.e., $\frac{4}{3}\pi R_h^3 \cdot \rho_s / \pi R_0^2 \cdot L \cdot \rho_s$, where $L$ is 60 nm for the case of irradiated rutile by 1390 MeV Bi ion) for each end of track.

Indeed, the "conical mass deficit" model is also applicable for the amorphizable materials. But, we must emphasize two things: First, the recrystallization process is negligible in amorphizable materials (leaving a cylindrical amorphous track) [65]. Second, it is very difficult to distinguish the small change near surface by TEM [69], which is covered by the background of amorphous cylindrical track.

Other phenomena in irradiated rutile can also be explained based on the above argument: (1) The radius of the thin track connected with two conical parts of track is about 1 nm (Fig. 1(d)). It coincides with the radius of core of thermal spike, wherein the duration of molten phase is the longest and the fluid velocity is highest. (2) The deduced radius by RBS/C is strikingly close to the calculated track size, i.e. the initial track size before recrystallization, by i-TS model. The recrystallized region from the initial track, which acts as defective channels around the final track, probably contributes to the backscattering yield for the RBS/C measurement [55,56].

## 5. Conclusion

We find the fine structure of latent tracks in rutile $TiO_2$ irradiated by swift heavy ions. Based on i-TS model, we give a new insight into the ion track formation in non-amorphizable materials. Actually, this model is applicable for one kind of materials that have a strong ability to recrystallize. Cylindrical, dumbbell shaped or sandglass-like track morphologies are formed due to outflow of the molten material and recrystallization as a function of ion path length. The specific angle and length of the cone depend on material and electronic energy loss.

## Acknowledgements

This work was financially supported by National Natural Science Foundation of China (Grant Nos. 11405229, 11690041, 41673062 and 11505243). P. Zhai acknowledges financial support of CAS "Light of West China" Program and the Natural Science Foundation of Gansu Province (Grant No. 18JR3RA392). W. Li acknowledges the support of the Thousand Young Talents Program of China. P. Zhai and J. Liu are grateful to Prof. Marcel Toulemonde for providing the inelastic thermal spike model code generously and some useful discussions. We would like to thank the accelerator staff of the HIRFL.

Giulian, M. C. Ridgway, A. P. Byrne, C. Trautmann, D. J. Cookson, K. Nordlund, and M. Toulemonde, Fine structure in swift heavy ion tracks in amorphous $SiO_2$, Phys. Rev. Lett. **101**, 175503 (2008).

**Table 1**. Irradiation parameters of rutile (mass density 4.25 g/cm$^3$) for ions with different incident energies. Electronic stopping power ($S_e$) and ion range ($R_p$) were calculated with SRIM2013 code [26]. Hillock height ($D_h$), apparent hillock/track diameter ($D_a$), corrected hillock diameter ($D_c$) and yield of track ($\eta$) were obtained from the measurements of AFM or TEM.

| Ion | Energy (MeV) | Incident angle (°) | $S_e$ (keV/nm) | $R_p$ (μm) | $D_h$ (nm) | $D_a$ (nm) | $D_c$ (nm) | $\eta$ |
|---|---|---|---|---|---|---|---|---|
| $^{129}$Xe | 860 | 0 | 26.9 | 36.6 | 4.0±1.2 | 18.7±7.1 | 4.4±1.7 | 0.91 |
| $^{129}$Xe | 1341 | 45 | 24.3 | 55.3 | 3.1±0.5 | 17.1±2.2 | 3.7±1.3 | 0.83 |
| $^{129}$Xe | 1641 | 0 | 22.5 | 68.1 | 2.4±0.6 | 14.1±3.0 | 2.5±1.3 | 0.90 |
| $^{129}$Xe | 1940 | 45 | 20.9 | 81.9 | 2.2±0.5 | 13.1±1.4 | 2.1±0.6 | 0.85 |
| $^{129}$Xe | 2194 | 45 | 19.7 | 94.4 | 1.8±0.3 | 11.3±1.4 | 1.6±0.3 | 0.92 |
| $^{129}$Xe | 2300 | 45 | 19.2 | 99.9 | 2.0±0.4 | 12.6±2.6 | 2.0±0.8 | 0.80 |
| $^{181}$Ta | 353 | 45 | 32.5 | 17.0 | 7.6±1.2 | | | |
| $^{181}$Ta | 829 | 7 | 35.1 | 30.8 | 7.6±0.8 | 25.0±2.0 | 7.8±1.0 | 0.97 |
| $^{181}$Ta | 1169 | 0 | 34.5 | 40.5 | 7.6±1.6 | | | |
| $^{181}$Ta | 1387 | 7 | 33.9 | 46.9 | 6.8±0.9 | 23.7±1.5 | 7.0±0.8 | 1.01 |
| $^{181}$Ta | 1703 | 7, 45 | 32.7 | 56.4 | 6.2±1.4 | 22.5±1.6 | 6.3±1.0 | 0.97 |
| $^{181}$Ta | 1786 | 7 | 32.4 | 58.7 | 6.3±0.8 | 22.1±1.4 | 6.1±0.8 | 1.03 |
| $^{181}$Ta | 1908 | 7 | 32.0 | 62.7 | 6.0±0.8 | 21.7±1.4 | 5.9±0.8 | 0.95 |
| $^{181}$Ta | 2329 | 0 | 30.4 | 76.2 | 5.5±0.9 | | | 1.10 |
| $^{209}$Bi | 894 | 60 | 40.1 | 29.7 | | 8.6±1.6 | | |
| $^{209}$Bi | 1390 | 0,60 | 39.9 | 42.1 | | 8.4±1.6 | | |

**Table 2**. Main physical parameters of rutile $TiO_2$ for i-TS model calculation.

|  | Rutile $TiO_2$ |
|---:|:---:|
| Density (g/cm$^3$) | 4.25 for solid, 3.21 for liquid [46] |
| Specific heat (J/(g K)) | 0.69 (300 K) - 0.99 (2100 K) [45] |
| Thermal conductivity (J/(cm K s)) | 0.104 (300 K) - 0.0321 (1400 K) [47] |
| Latent heat of fusion (J/g) | 838.07 [45] |
| Melting point (K) | 2130 [45] |
| Energy to melt (eV/at) | 0.69 [45] |
| Band gap (eV) | 3.05 [48] |
| Electron-phonon mean free path (nm) | 5.8 [19,41-44] |

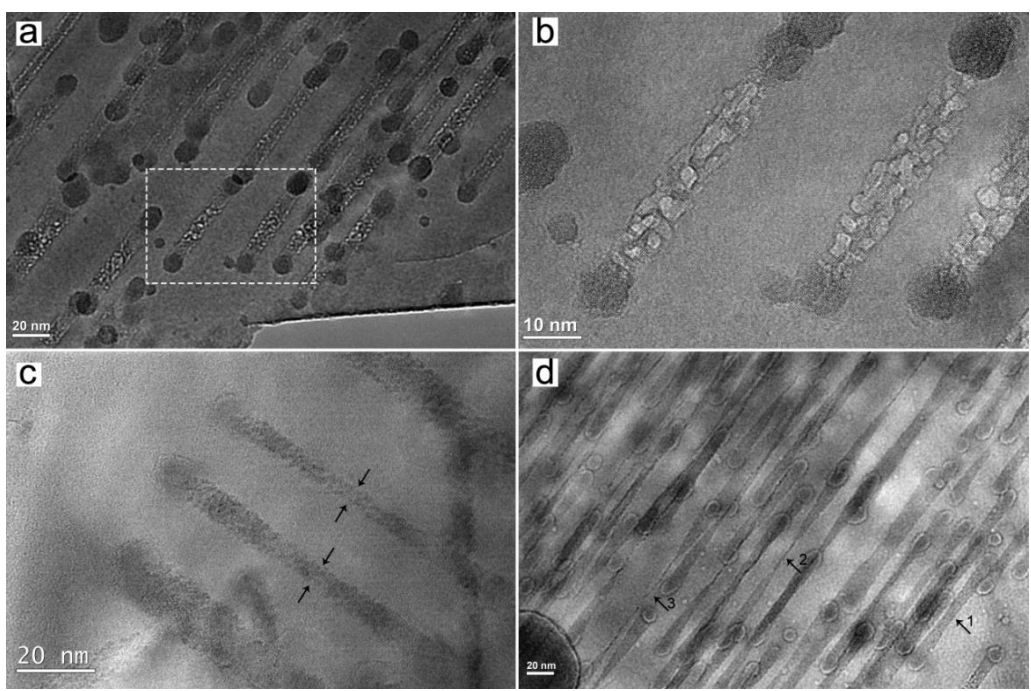

FIG. 1. Bright field TEM micrographs of ion tracks with different morphologies induced by 1390 MeV Bi ions in lamella rutile. (a) Ion tracks with nearly cylindrical morphology (underfocused condition). A pair of spherical hillocks at each end of an ion track can be found on both the top and bottom surface when the energetic ions penetrate the entire thickness of a thin sample. (b) HRTEM micrograph from the dotted square in (a) evidently shows that an ion track consists of many irregular shaped voids (underfocused condition). (c) Two dumbbell-shaped ion tracks where the "waist" locations are marked by arrows are visible. (d) Ion track morphologies are strongly correlated with the actual ion path length in the sample (three morphologies marked by arrows can be identified).

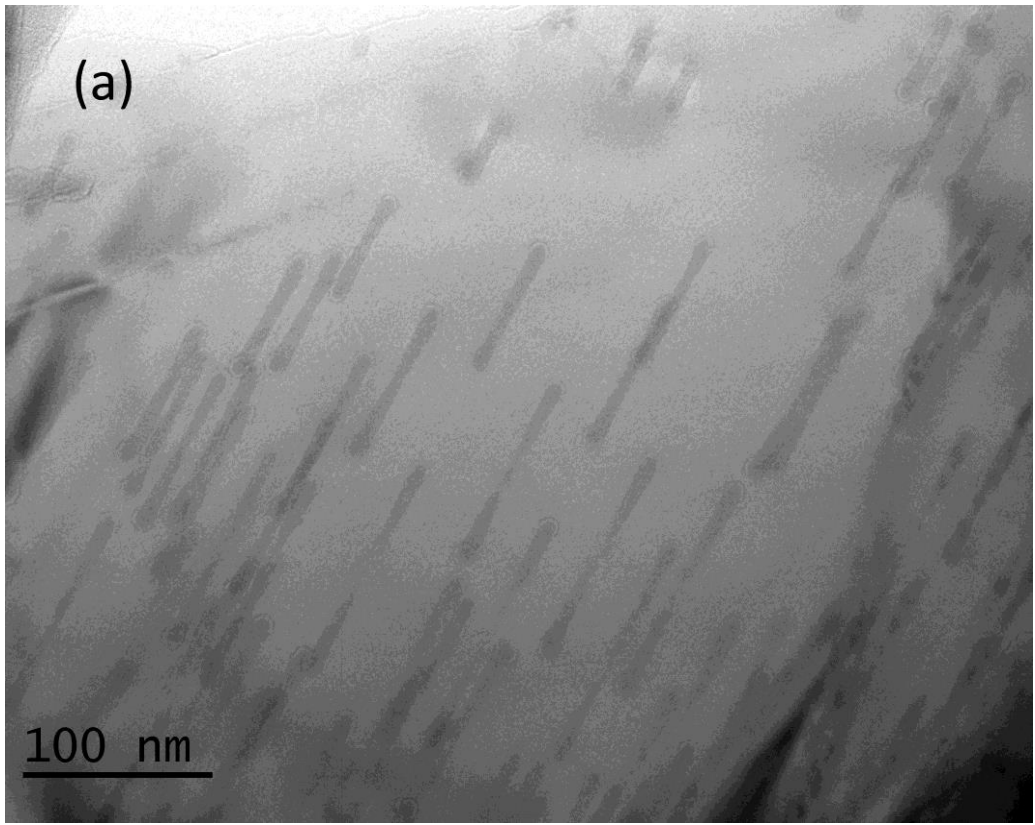

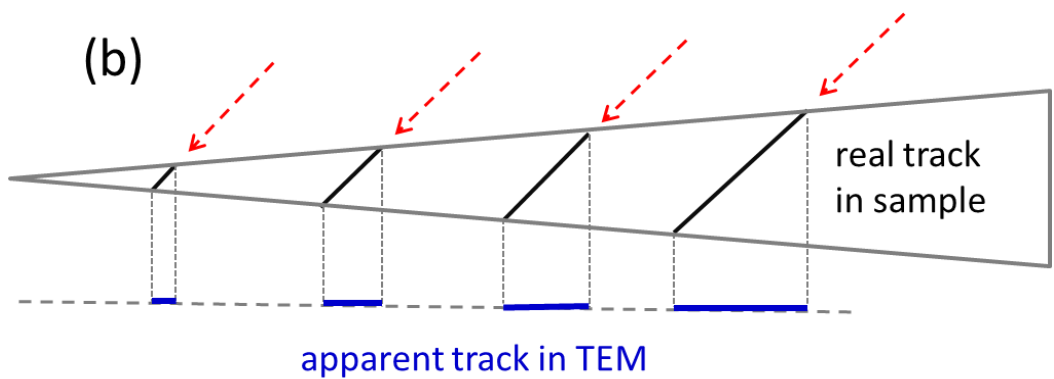

FIG. 2. Bright field TEM micrographs (a) of ion tracks with different morphologies in lamella rutile induced by 1703 MeV Ta ions with 45$^\circ$ incidence, and schematic (b) for the positive relation between real track in sample and apparent track in TEM.

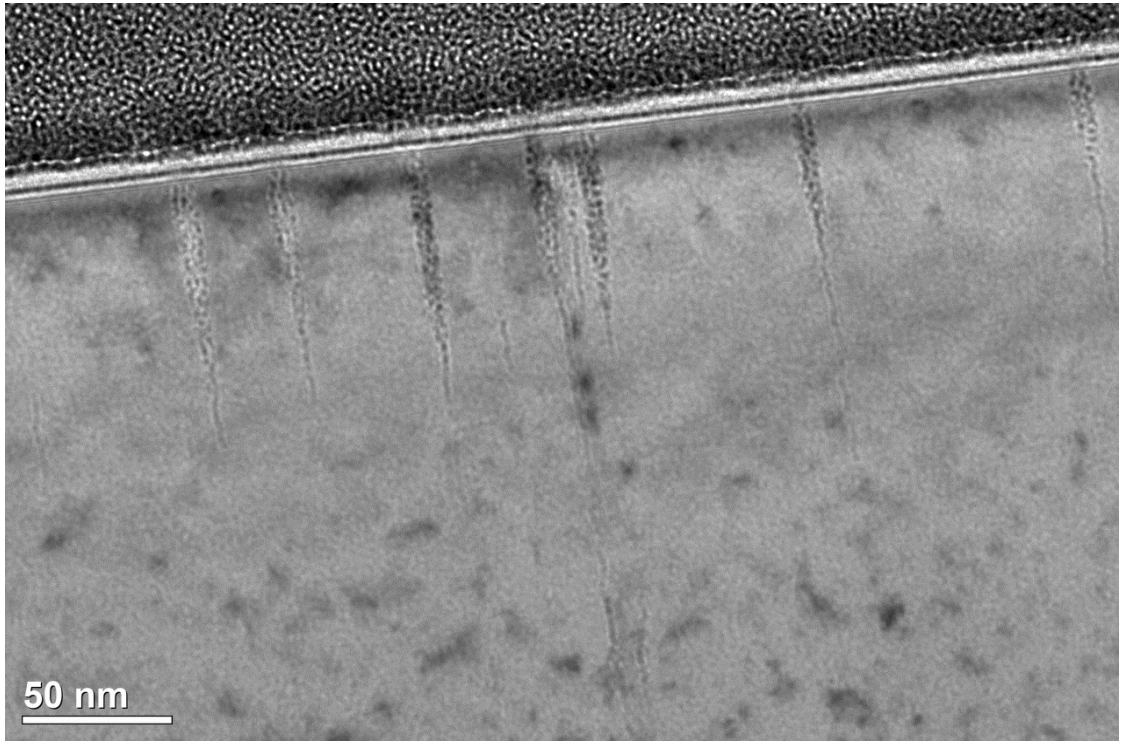

FIG. 3. Cross-sectional TEM image of the conical ion track in bulk rutile $TiO_2$ irradiated by 1390 MeV Bi ions.

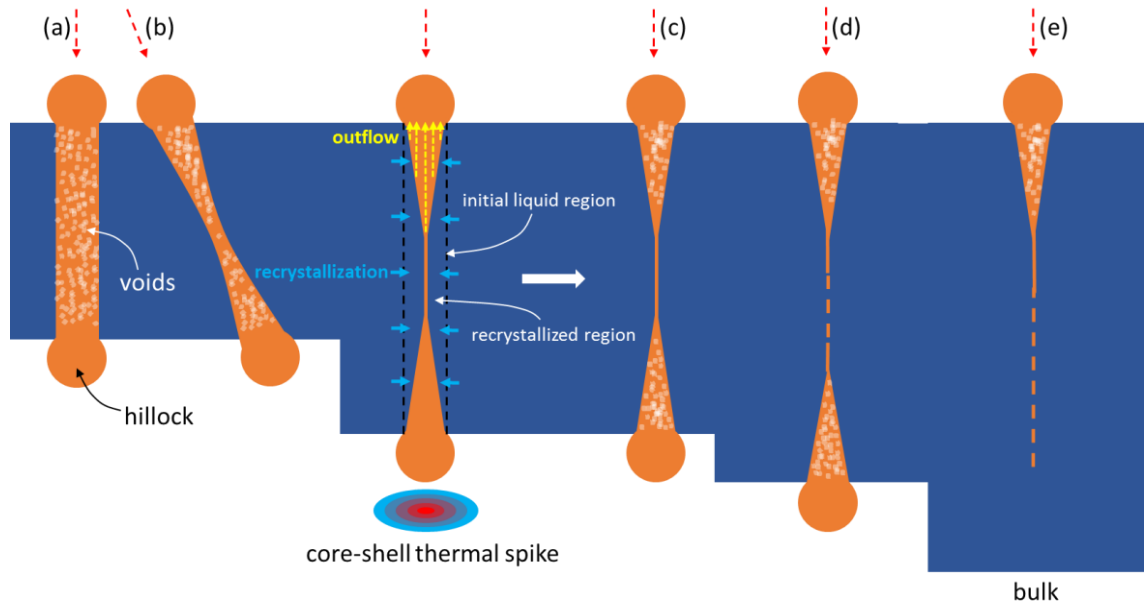

FIG. 4. Schematic for ion track morphology evolution as a function of ion path length in rutile. In the case of irradiated rutile by 1390 MeV Bi ions, (a) When the ion path length is shorter than about 60 nm, the track morphology is nearly cylindrical. (b) When the ion path length is longer than 60 nm, the dumbbell-shaped morphology emerges. (c) When the ion path length is about 150 nm, two conical features near surface connected with a thin track appears. (d) When the ion path length is longer than 150 nm, the thin track connecting two conical features near surface start to be discontinuous. (e) For the bulk rutile, the conical track appears near the irradiated surface. In the deep, the thin track is discontinuous and finally disappears. In our model, the core-shell structure of thermal spike, liquid outflow and recrystallization contribute to the morphology evolution with the ion path length.

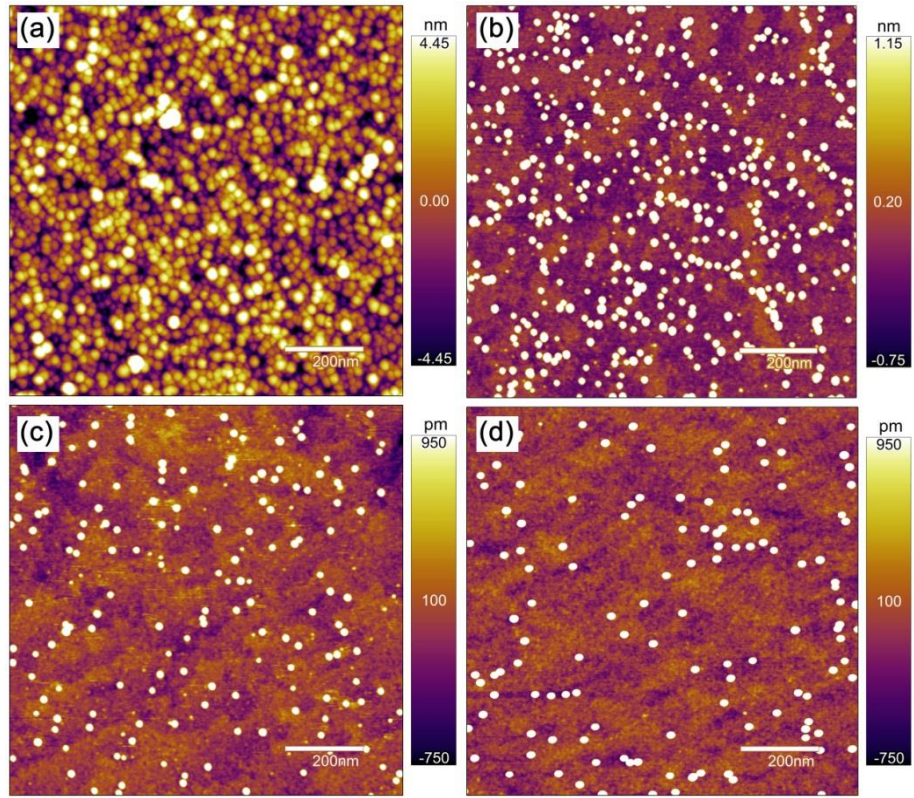

FIG. 5. AFM micrograph of hillocks on the surface of rutile irradiated with swift heavy ions. (a) Hillocks induced by 1390 MeV Bi ions ($5\times10^{11}$ ions/cm$^2$) appear overlapped due to the finite curvature radius of the AFM tip. (b) Hillocks induced by 860 MeV Xe ions ($5\times10^{10}$ ions/cm$^2$). (c) Hillocks induced by 1341 MeV Xe ions ($2\times10^{10}$ ions/cm$^2$). (d) Hillocks induced by 1908 MeV Ta ions ($1\times10^{10}$ ions/cm$^2$). The yield of track is about 80-100%, implying that one ion creates one track.

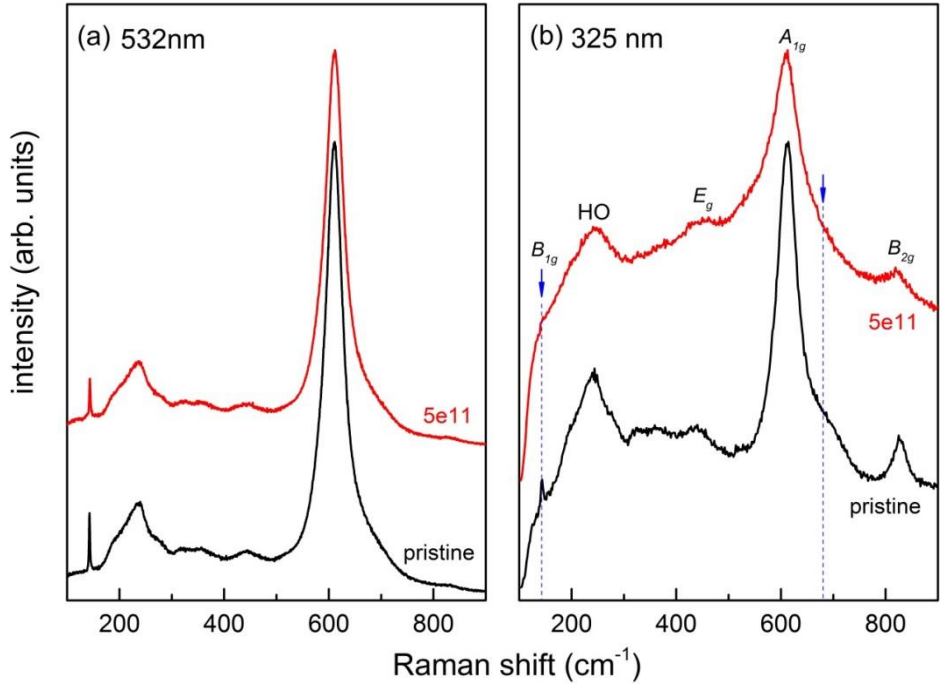

FIG. 6. Raman spectra of bulk rutile (001) irradiated by 1390 MeV Bi ions at a fluence of $5 \times 10^{11}$ ions/cm$^2$. (a) Visible Raman spectra of irradiated sample excited by 532 nm light with the sampled depth about 2 μm, shows no significant change. (b) UV Raman spectra of irradiated sample, excited by 325 nm light with the detection depth about 20 nm, reveals the disappearance of B$_{1g}$ (143 cm$^{-1}$) and the shoulder (~700 cm$^{-1}$) of A$_{1g}$ bands.

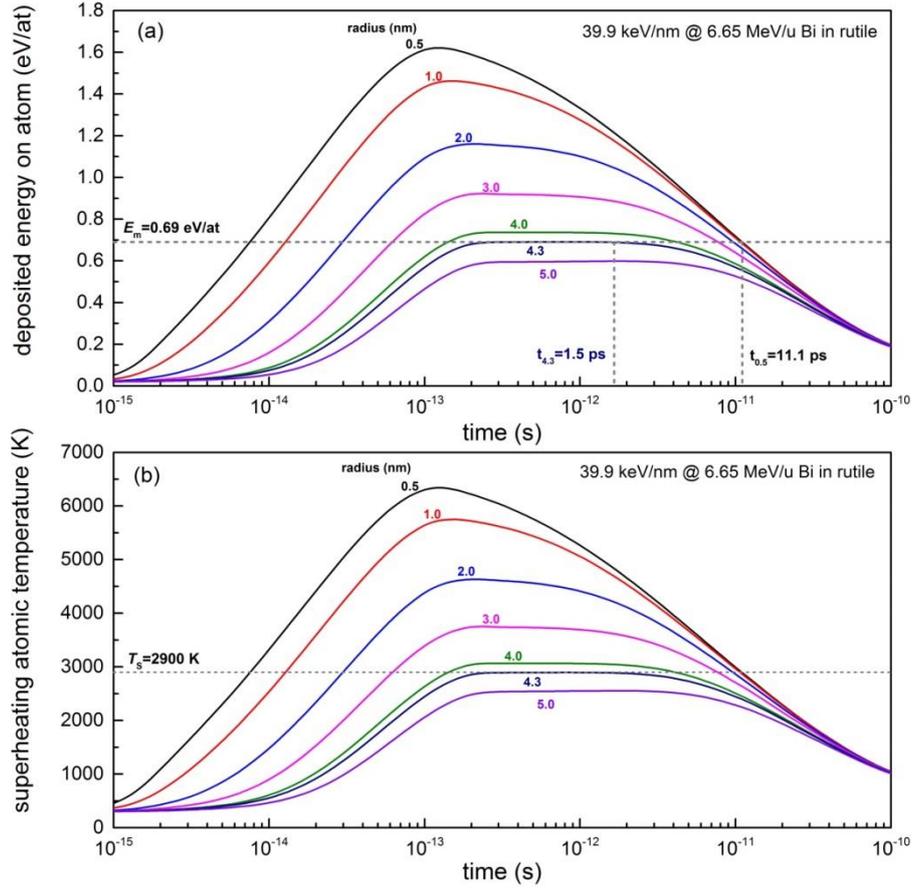

FIG. 7. i-TS calculations for 1390 MeV Bi irradiation in rutile. (a) Deposited energy on atoms versus time. Using the energy to melt $E_m$ 0.69 eV/at [45], the deduced radius is 4.3 nm. The duration of molten phase for the radius of 0.5 and 4.3 nm are 11.1 and 1.5 ps, respectively. (b) Superheating atomic temperature versus time. For the radius of 4.3 nm, superheating melting temperature is 2900 K.

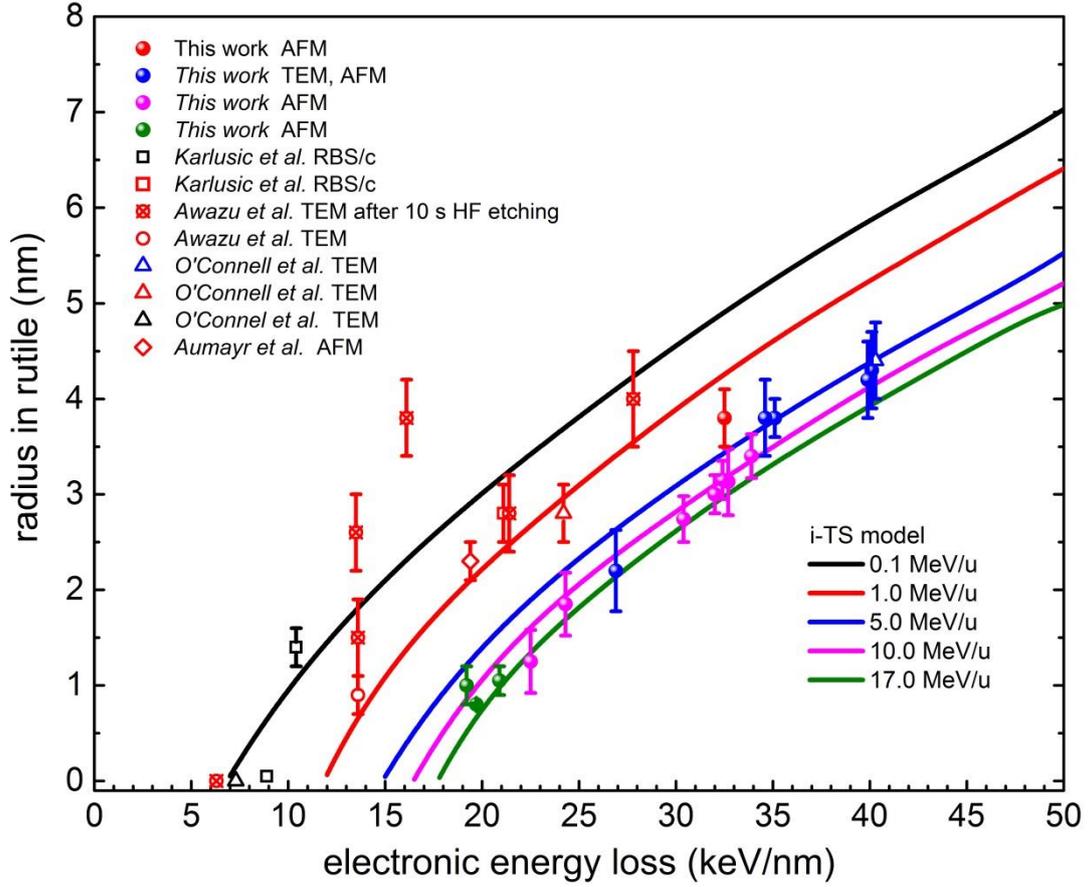

FIG. 8. Experimental radii of ion track in single crystal rutile versus electronic energy loss. The experimental results (with five kinds of colors correspond to different specific energies) are reproduced from Karlusic et al. [23], Awazu et al. [19,53], O'Connell et al. [18], Aumayr et al. [54] and this work. The solid lines are i-TS calculations for five different specific energies of 0.1, 1.0, 5.0, 10.0 and 17.0 MeV/u, respectively. Note that the data (open circle with cross) reproduced from Awazu et al. [19] are determined by TEM after 20% hydrofluoric acid etching for 10 seconds.